\documentstyle[12pt]{article}
\newcommand{\be}{\begin{equation}}
\newcommand{\ee}{\end{equation}}

\newcommand{\go}{\omega}
\newcommand{\half}{\frac{1}{2}}
\newcommand{\ga}{\alpha}

\newcommand{\gb}{\beta}
\newcommand{\gee}{\epsilon}
\newcommand{\ggg}{\gamma}

\newcommand{\bee}{\begin{eqnarray}}
\newcommand{\eee}{\end{eqnarray}}
\newcommand{\nn}{\nonumber}
\begin{document}
\begin{flushright}
 FIAN/TD/27--97\\
\end{flushright}\vspace{2cm}

\begin{center}
{\large\bf DEFORMED OSCILLATOR ALGEBRAS AND
HIGHER-SPIN GAUGE INTERACTIONS OF MATTER FIELDS IN 2+1 DIMENSIONS}
\vglue 0.6  true cm
\vskip1cm
{\bf M.A.~VASILIEV}
\vskip0.5cm
I.E.Tamm Department of Theoretical Physics, Lebedev Physical Institute,\\
Leninsky prospect 53, 117924, Moscow, Russia
\vskip0.5cm
\vglue 0.3  true cm
\end{center}
\vskip0.8cm
\begin{abstract}
We formulate a non-linear system of equations which describe
higher-spin gauge interactions of massive matter fields
in 2+1  dimensional space-time and explain some properties
of the deformed oscillator algebra which underlies this
formulation. In particular we show that the
parameter of mass $M$ of matter fields is related to the
deformation parameter in this algebra.
\end{abstract}

\section{Introduction}

Dmitrij Vassilievich Volkov was a brilliant scientist
who had made a great
contribution to theoretical physics
and created a remarkable scientific school in Kharkov.
His most famous results are related to the
creation of supersymmetric theories. The original approach
invented by  Volkov and collaborators was based on the
application of invariant connection forms \cite{v}. A further
development of these  geometric ideas in
the modern field theory was extremely fruitful.
In this talk we argue that a proper
generalization of the Volkov's ideas leads to a universal
method of description of relativistic dynamics in terms of certain
zero-curvature conditions supplemented with appropriate constraints. We will
illustrate this by considering an example of matter fields
interacting through higher-spin (HS) gauge fields
in 2+1 dimensions.

\section{Higher-Spin Symmetries in 2+1 Dimensions and
Deformed Oscillator Algebras}

HS algebras  in $d$ space-time dimensions
are certain infinite-dimensional extensions of
space-time symmetry algebras $s_d$
\cite{FV,V1}, which act on appropriate physical fields.
HS symmetries can be gauged by virtue of introducing appropriate
HS gauge fields.
In 2+1 dimensions, HS gauge fields do not propagate rather mediating
interactions of
matter sources analogously to the case of the gravitational field in 2+1
dimensions \cite{T,W}. This is a greatly simplifying property
compared to the HS dynamics in four and higher dimensions.
HS symmetries in 2+1 dimensions  are still non-trivial
as well as HS matter multiplets, i.e. the multiplets of fields on which
the HS symmetries are realized. They are
however very simple:  ordinary scalar
and spinor fields of an arbitrary mass.
The analysis of HS interactions of  relatively simple lower dimensional
models sheds some light on general properties of HS models.

It is most useful to start with
the space-time symmetries of (anti) - de Sitter type $s_d = o(d-1,2)$
analyzing a possibility of taking a flat limit afterwards.
The case of $d=3$ is
special because $s_3 =o(1,2)\oplus o(1,2)=sp(2)\oplus sp(2) $ is not
simple.
Originally it was conjectured \cite{bl} that
a 3d HS algebra is the direct sum
of two Heisenberg-Weyl algebras (more precisely, of
their Lie supercommutator
superalgebras), each constructed from the ordinary oscillators
 \footnote{
 The indices $\ga , \,\gb\,,\ggg\, =1,2$ are treated as spinor indices
 in 2+1 dimensions. These are lowered and raised with the
 aid of  the symplectic form $\gee_{\ga\gb}=-\gee_{\gb\ga}$,
 $\gee_{ 12} =\gee^{12}=1$,
 $A^\ga =\gee^{\ga\gb}A_\gb$,
 $A_\ga =
 A^\gb
 \gee_{\gb\ga}$.}
$[y_\alpha ,y_\beta ] = 2i\epsilon_{\alpha\beta}$.
Because 3d HS gauge fields are not propagating one can write
the Chern-Simons action for the pure gauge
HS system,
$
S=\int_{M_3}str (A\wedge dA +\frac{2}{3}A\wedge A\wedge A )\,
$
with the gauge fields $A$ taking values in the HS algebra.
In \cite{BBS,H,Q} it was shown that there exists a one-parametric class
of infinite-dimensional algebras which we denote $ hs(2;\nu )$
($\nu$ is an arbitrary real parameter), all containing $sp(2)$ as a
subalgebra.
This allows one to define a class of HS algebras
$
g=hs(2;\nu )\oplus hs(2;\nu ).
$
The supertrace operation
was defined in \cite{Q} where also a
useful realization of
the supersymmetric extension of
$ hs(2;\nu )$ was given, based on a certain deformed oscillator algebra.
Since this construction will be used below and also
gets interesting applications in a number
of different physical problems let us explain its properties in
somewhat more details.

Consider an associative algebra
$Aq(2;\nu )$
with a general element of the form
\be
\label{sel}
f(q,K )=
\sum^\infty_{n=0}\sum_{A=0,1}
\!
\frac{1}{n!}
 f^{A\,\alpha_1\ldots\alpha_n}(K)^A
 q_{\alpha_1}\ldots q_{\alpha_n}\,,
\ee
under condition that the coefficients
$ f^{A\,\alpha_1\ldots\alpha_n}$ are symmetric with respect to
 the indices
$\ga_j =1,2$ and that the generating elements $q_\ga$ satisfy the
 relations

\bee
\label{modosc}
[q_\ga ,q_\gb ]=2i
\gee_{\ga\gb} (1+\nu K)\,,\quad Kq_\ga
=-q_\ga K\,,
\quad K^2 =1\,,
\eee where $\nu $ is an arbitrary parameter.
In other words,
$Aq(2;\nu )$ is the enveloping algebra for the relations
(\ref{modosc}), the deformed oscillator algebra.

An important property of this algebra is that for all $\nu$
the bilinears
\be
\label{lorq}
T_{\ga\gb}=\frac{1}{4i}\{q_\ga \,,q_\gb \}
\ee
have $sp(2)$ commutation relations and rotate $q_\ga$ as a $sp(2)$ vector
\bee
\label{q1com}
[T_{\ga \gb},
T_{\ggg \eta}]\!=\!( \gee_{\ga\ggg} T_{\gb\eta}\!+\!
\gee_{\gb\ggg} T_{\ga\eta}\!+\!
\gee_{\ga\eta} T_{\gb\ggg}\!+\! \gee_{\gb\eta} T_{\ga\ggg}),\quad\!\!\!
\eee
\bee
\label{q2com}
[T_{\ga\gb} , q_{\ggg }]
\!=\! \gee_{\ga\ggg}q_{\gb} \!+\! \gee_{\gb\ggg}q_{\ga}.
\eee

The deformed oscillators described above
have a long history and
were originally discovered by Wigner \cite{wig}
who addressed a question whether it is possible to modify
the commutation relations for the normal oscillators
$a^\pm$ in such a way
that the basic commutation relations $[H, a^\pm] =\pm a^\pm $,
$H=\frac{1}{2} \{a^+ ,a^- \}$ remain
valid. By analyzing this problem in the Fock-type space
Wigner found a one-parametric deformation of the standard commutation
relations which corresponds to a particular realization of the
commutation relations (\ref{modosc})  with the identification
$a^+ =q_1$,
$a^- =\frac{1}{2i} q_2$, $H=T_{01}$ and $K=(-1)^N$ where $N$ is the
particle number operator. These commutation relations were
discussed later by various authors in particular
in the context of parastatistics (see e.g. \cite{des}).

According to (\ref{lorq}) and (\ref{q2com})
the $sp(2)$ symmetry generated by $T_{\ga\gb}$
extends to $osp(1,2)$
supersymmetry by identifying the supergenerators with
$q_\ga$. In fact, as shown in \cite{BWV}, one can start from the
$osp(1,2)$ algebra to derive the deformed oscillator commutation
relations. Since this construction is instructive in many respects
we reproduce it here.

One starts with the (super)generators
$T_{\alpha\beta}$ and $q_\alpha$
which by definition of $osp(1,2)$ satisfy the commutation
relations (\ref{lorq})-(\ref{q2com}). Since
 $\alpha$ and $\beta$ take only two values one can write
\be
\label{co}
[q_\ga ,q_\gb ]=2i
\gee_{\ga\gb} (1+Q)\,,
\ee
where $Q$ is some new ``operator" while the unit term is singled out for
convenience.
Inserting this
back into (\ref{q2com}) with the substitution
of (\ref{lorq}) and completing the commutations
one observes that (\ref{q2com}) is
true if and only if $Q$ anticommutes with $q_\alpha$\,,
\be
\label{Q}
Q q_\alpha =-q_\ga Q.
\ee
The relation (\ref{q1com}) does not add anything new since it
is a consequence of (\ref{lorq}) and (\ref{q2com}).
As a result we arrive \cite{BWV} at the following
important

{\bf Corollary}: {\it The enveloping algebra of $osp(1,2)$,
$U(osp(1,2))$, is isomorphic to the enveloping algebra of the deformed
oscillator relations (\ref{co}) and (\ref{Q}).}

In other words, the associative algebra with the generating elements
$q_\ga$ and $Q$ subject to the
relations (\ref{co}) and (\ref{Q})
is the same as the associative algebra with the generating elements
$q_\ga$ and $T_{\ga\gb}$ subject to the osp(1,2) commutation relations
(\ref{lorq})-(\ref{q2com}).

Computing the quadratic Casimir operator of $osp(1,2)$
\be
C_2 = -\frac{1}{2}T_{\ga\gb}T^{\ga\gb} -\frac{i}{4}q_\ga q^\ga.
\ee
one easily derives using (\ref{co}) that
\be
C_2 =-
\frac{1}{4} (1- Q^2)\,.
\ee

Let us now consider the factor algebra of $U(osp(1,2))$ over its
ideal
$I_{(C_2 + \frac{1}{4} (1- \nu^2))}$
generated by the element
$(C_2 + \frac{1}{4} (1- \nu^2) )$
where $\nu$ is an
arbitrary number. In other words we assume that every element of
$U(osp(1,2))$ which is of the form
$\left( C_2 + \frac{1}{4} (1- \nu^2)\right) a$,
$\forall a\in U(osp(1,2))$ is equivalent to zero.
This factorization can be achieved in terms of the
deformed oscillators (\ref{co}), (\ref{Q}) by setting
\footnote{The point $\nu =0$ is special since one can
consider a case(s) with $Q^2 =0$, $Q\neq 0$.}
\be
Q=\nu K\,,\qquad K^2 =1 \qquad Kq_\ga =-q_\ga K\,.
\ee

Thus, it is shown \cite{BWV} that the algebra $Aq(2,\nu )$ introduced in
\cite{Q} is isomorphic to
$U(osp(1,2))/
I_{(C_2 + \frac{1}{4} (1- \nu^2))}$.
This fact has  a number of simple
but important consequences. For example, any representation of the
superalgebra $osp(1,2)$ with
$C_2 =  -\frac{1}{4} (1- \nu^2)$
forms a representation of
$Aq(2,\nu )$ ($\nu \neq 0$) and vice versa (for all $\nu$ including
$\nu = 0$).
In particular this is the case for finite-dimensional
representations corresponding to the values $\nu =2l+1$, $l\in
{\bf Z}$ with $C_2 = l(l+1)$. This fact has been used in \cite{br} for the
construction of the generalized Toda field theories interpolating between
ordinary finite-component Toda field theories. Let us note that the even
subalgebra of $Aq(2;\nu )$
spanned by the elements of the form (\ref{sel}) with
$f(q,K)=f(-q,K)$
decomposes into a direct sum of two subalgebras
$Aq^E_\pm (2;\nu )$ spanned by the elements
$P_\pm f(q,K)$ with
$f(-q,K)=f(q,K)$, $P_\pm = \frac{1}{2} (1\pm K)$. These algebras can be
shown  to be  isomorphic to the factor algebras
$U(sp(2))/I_{(C_2 + \frac{3\pm 2\nu -\nu^2}{4})} $
where
$C_2 = -\frac{1}{2}T_{\ga\gb}T^{\ga\gb}$ is the quadratic
Casimir operator of $sp(2)$ and can be interpreted
as (infinite-dimensional) algebras interpolating between the ordinary
finite-dimensional matrix algebras. Such interpretation
of $U(sp(2))/I_{(C_2 -c)} $
was given by Feigin in \cite{BF}.

A very important property of
$Aq(2;\nu )$
is that it admits \cite{Q} a uniquely defined supertrace operation
\be
\label{str}
str (f) =f^0 -\nu f^1\,,
\ee
such that
$str(fg) = (-1)^{\pi_f \pi_g} str(gf)
$, $\forall f,g $ having a definite parity,
$f(-q,K)
\!\!=\!\!(-1)^{\pi_f}\!f(q,K)
$
(i.e. $str (1)=1\,,\quad str(K) =-\nu$ while all higher
 monomials of $q_\ga$ in (\ref{sel}) do not contribute under the supertrace).
This supertrace reduces \cite{br}
to the ordinary supertrace of finite-dimensional algebras
for the special values of the parameter $\nu=2l+1$ which correspond
to the values of the Casimir operator related to finite-dimensional
representations of $osp(1,2)$ ($sp(2)$ in the bosonic case).
This property allows one to handle the algebras
$Aq(2;\nu )$ very much the same way as ordinary finite-dimensional
(super)matrix algebras. What happens for special values of
$\nu=2l+1$ is that
$Aq(2;\nu )$
acquires ideals $I_l$ such that
$Aq(2;\nu )/I_l$
amounts to appropriate (super)matrix algebras. These ideals
were described  in \cite{Q} as null vectors of the invariant
bilinear form $str (ab)$, $a,b \in Aq(2;\nu )$.

The identification  of
$Aq(2;\nu )$ with
$U(osp(1,2))/
I_{(C_2 + \frac{1}{4} (1- \nu^2))}$
makes transparent such properties of
the deformed oscillator algebra as relationship of the representations
of
$Aq(2;\nu )$ with those of $osp(1,2)$ (including its finite-dimensional
representations for special values of $\nu=2l+1, \forall l \in {\bf Z} $)
and $N=1$ supersymmetry (as inner $osp(1,2)$ automorphisms). A more
interesting property \cite{BWV} is that
$Aq(2;\nu )$ admits $N=2$ supersymmetry
$osp(2,2)$ with the generators
\be
\label{N2}
T_{\ga\gb}=\frac{1}{4i}\{q_\ga \,,q_\gb \}\,,\quad
Q_\ga =q_\ga \,,\nn\\
S_\ga =q_\ga K\,,\quad J=K+\nu\,.
\ee
These properties find interesting applications
(see, e.g., \cite{plyu} and references therein).

As we demonstrate  below the deformed oscillator algebras serve as a
main tool for the description of the d=3 HS dynamics. The reason is
 that they are related to the enveloping algebras of the
space-time symmetries  and allow us to
formulate a non-linear dynamics with explicit local Lorentz symmetry.
In its turn, the analysis of the HS dynamics presented
below is interesting in the context of the deformed oscillator algebra
itself because algebraically it reduces to the construction of its
embedding  into a direct product
of two ordinary Heisenberg-Weyl (i.e. oscillator) algebras equipped
with certain twist operators.

Coming back to the HS problem in 2+1 dimensions we note that
to describe a doubling of the elementary algebras in
$g=hs(2;\nu )\oplus hs(2;\nu )$ it suffices to
introduce an additional central involutive
 generating element $\psi$:
$[\psi ,q_\ga ]=0$, $ [\psi ,K]=0,$
$\psi^2 =1.$
 The two simple subalgebras
of $g$ are singled out by the projection operators
$\Pi_\pm$ = $\half (1\pm \psi )$.
The full set of HS gauge fields in 2+1 dimensions,
the gauge fields of $g$, thus is
\bee
\label{gau}
A(q,K,\psi |x)\!\!=\!\!dx^\nu \sum_{n=0}^\infty\sum_{B=0,1}
\!\!\frac{1}{2i\,n!} (\go_\nu^{B\ga_1 \ldots \ga_n }(x)
\!+\!\psi
h_\nu^{B\ga_1 \ldots \ga_n }(x))\,(K)^B\,
q_{\ga_1}\ldots q_{\ga_n}.
\eee

The field strengths and gauge transformation laws are defined
in the usual way
\bee
R(q,K,\psi |x)=dA(q,K,\psi |x)
+A(q,K,\psi |x)\wedge
A(q,K,\psi |x)\,,\\
\delta A(q,K,\psi |x)=d\gee(q,K,\psi |x)
+[A(q,K,\psi |x)\,,
\gee (q,K,\psi |x)]\,,
\eee
where
$
d=dx^\nu
\frac{\partial}{\partial x^\nu }\,.
$
The gravitational fields
\be
A^{gr} =\frac{1}{4i} (\go^{\ga\gb} +h^{\ga\gb}\psi )q_\ga q_\gb
\ee
take values in the subalgebra $sp(2)\oplus sp(2)$.
The pure gauge  Chern-Simons
HS action reduces to the Witten gravity action
\cite{W} in the
spin 2 sector and to the Blencowe's HS action \cite{bl}
in the case of $\nu =0$.

\section{Unfolded equations}
In this report which is based on a recent papers
\cite{pr1,pr2} we answer to the
question how to introduce
interactions of HS gauge fields with propagating matter fields
at the level of equations of motion using an approach
 which we call ``unfolded formulation" \cite{un}.
It consists of reformulation of  dynamical equations in a form of certain
 zero-curvature conditions and covariant constancy
 conditions
\be
\label{0cur}
d\go +\go\wedge\go=0\,,\qquad
dB^A +\go^i t_i \, {}^A{}_B B^B=0\,,
\ee
supplemented with some gauge invariant constraints
\be
\label{con}
\chi (B) =0 \,
\ee
which do not contain
space-time derivatives.
Here
 $\go (x)=dx^\nu \go_\nu^i (x) T_i $ is a gauge field
taking values in some Lie superalgebra $l$ ($T_i \in l)$,
and $B^A (x)$ is a set of 0-forms  which take values in
the representation space of
some representation $(t_i ){}^B {}_A$ of $l$.

An interesting property of this form of
equations is that their
dynamical content
is hidden  in the constraints (\ref{con}).
Indeed, locally one can integrate out
(\ref{0cur}) explicitly as
$
\go= g (x)d g^{-1} (x),
$
$
B(x)=t_{g(x)} (B_0 )
$
where $g(x)$ is an arbitrary invertible element
while
$B_0$ is an arbitrary $x$ - independent
representation element and
$t_{g(x)}$ is the exponential of the representation $t$ of $l$.
Since the constraints $\chi (B)$ are gauge invariant one is left with the
only condition
$
\chi (B_0 )=0\,.$
Let $g(x_0 )=I$ for some point of space-time  $x_0$.
Then $B_0$=$B(x_0 )$.

Such a formulation can in principle be applied to an arbitrary
dynamical system provided that a representation $t$ is infinite-dimensional.
Being based on zero-curvature conditions it has deep similarities with
the original approach by Volkov and collaborators \cite{v}.
A crucial feature of our approach is that
 the set of 0-forms $B$ has to be reach enough to describe
all space-time derivatives of the dynamical fields
while the constraints (\ref{con}) effectively impose all
restrictions on the space-time derivatives required by the
dynamical equations under consideration.
Given solution
of (\ref{con}) one knows all derivatives of the dynamical fields
 compatible with the field equations and can therefore reconstruct
these fields by analyticity in some neighborhood of $x_0$.
The specificity of the HS dynamics
which makes such an approach adequate is that HS
 symmetries mix all orders of derivatives which therefore
 have to be contained in a
 representation $t$  of
 HS symmetries.

Let us illustrate this by the example
 of a scalar field $\phi$ obeying the massless Klein-Gordon
 equation $\Box\phi=0$ in a flat  space-time
 of an arbitrary dimension $d$.
 Here $l$ is identified with
the Poincare algebra $iso(d-1,1)$ which gives rise to the gauge fields
$
\go_\nu =(h_\nu{}^a ,\go_\nu {}^{ab} )
$ ($a,b =0-(d-1)$).
The zero curvature conditions of $iso(d-1,1)$,
$R_{\nu\mu}{}^a =0$ and $ R_{\nu\mu}{}^{ab}=0,$
imply that the vierbein $h_\nu {}^a$
and Lorentz connection $\go_\nu {}^{ab}$ describe the flat geometry.
Fixing the local Poincare gauge transformations one can set
\be
\label{flgauge}
h_\nu {}^a =\delta _\nu^a \,,\quad \go_\nu{}^{ab}=0\,.
\ee

To describe dynamics of a spin zero massless field
$\phi (x)$ let us introduce an
infinite collection of 0-forms $\phi_{a_1\ldots a_n}(x)$
which are totally symmetric traceless tensors
\be
\label{tr}
\eta^{bc}\phi_{bca_3\ldots a_n}=0\,,
\ee
where $\eta^{bc}$ is the flat Minkowski metrics.
The ``unfolded" version of the Klein-Gordon equation
has a form of the following infinite chain of equations
\be
\label{un0}
\partial_\nu \phi_{a_1\ldots a_n }(x) =h_\nu {}^b
\phi_{a_1 \ldots a_n b}(x)\,,
\ee
where we have replaced the Lorentz covariant
derivative by the ordinary flat derivative $\partial_\nu$ using the
gauge condition (\ref{flgauge}).
The tracelessness condition (\ref{tr})  is a specific realization of the
constraints (\ref{con}) while the system of equations
(\ref{un0}) is a particular example of the equations (\ref{0cur}).
It is easy to see that this system is formally
consistent.

To show that the system (\ref{un0}) is equivalent to the
free massless field equation $\Box \phi (x)=0$ let us identify the
scalar field $\phi (x)$ with the $n=0$ member of the tower
of 0-forms $\phi_{a_1 \ldots a_n}$.
Then the first two equations
(\ref{un0}) read
$
\partial_\nu \phi =\phi_\nu
$
and
$
\partial_\nu \phi_\mu= \phi_{\mu\nu}\,,
$
respectively.
The former tells us that
$\phi_\nu$ is a first derivative of $\phi$.
The latter implies that
$\phi_{\nu\mu}$ is a second derivative of $\phi$. However, because of the
tracelessness condition (\ref{tr}) it imposes the Klein-Gordon
equation
$\Box\phi =0$.
It is easy to see that all other equations in (\ref{un0}) express highest
tensors in terms of the higher-order derivatives
$
\label{hder}
\phi_{\nu_1 \ldots \nu_n}= \partial_{\nu_1}\ldots\partial_{\nu_n}\phi
$
and impose no additional conditions on $\phi$. The tracelessness conditions
are all satisfied once the Klein-Gordon equation is true.

\section{Free fields in 2+1 AdS space}
Let us now confine ourselves to the 2+1 dimensional
case and generalize the above analysis of the scalar
field dynamics to the AdS geometry. The gauge fields
of the AdS algebra $o(2,2)\sim sp(2)\oplus sp(2)$
are identified with the gravitational
fields, $A_\nu =(\lambda h_{\nu \,,\ga\gb}\,;\go_{\nu \,,\ga\gb} )$.
The zero-curvature conditions $R_{\nu\mu}=0$ for the AdS algebra
in its orthogonal realization take a form
\be
\label{00cur}
R_{\nu\mu\,,ab}=\lambda^2 (h_{\nu a}h_{\mu b} -h_{\nu b}h_{\mu a})\,,\quad
R_{\nu\mu\,,a}=0
\ee
$ (\nu\,,\mu\ldots ; a\,,b\ldots =0-2)$,
where $R_{\nu\mu\,,ab}$ and $R_{\nu\mu\,,a}$ are the
Riemann and torsion tensors, respectively. {}From (\ref{00cur})
one concludes that the zero curvature
equations for the algebra $o(2,2)$ on a 3d
manifold do indeed describe the AdS space provided that
$h_\nu {}^a$ is identified with a dreibein and is invertible.

It is an important property of the 3d geometry
 that one can resolve the tracelessness conditions (\ref{tr})
by using the formalism of two-component
spinors: a totally symmetric traceless tensor $\phi _{a_1 \ldots a_n}$
is equivalent to a totally symmetric multispinor
$C_{\ga_1 \ldots \ga_{2n}}$. Let us now address the question
what is a general form of the
 equations analogous to (\ref{un0})
such that their   integrability conditions reduce to (\ref{00cur}).
The result is  that, up to a freedom in field redefinitions,
these are equations of the
form \cite{un}
\bee
D C_{\ga_1 \ldots \ga_{2n}} =h^{\gb\gamma}C_{\ga_1 \ldots\ga_{2n}
\gb\gamma}
\label{unads}
+2n(2n-1) e(2n,\lambda,M)h_{\{ \ga_1\ga_2} C_{\ga_3 \ldots
\ga_{2n}\}_\ga}\,,
\eee
where $D$ is the Lorentz covariant derivative,
$
D B_\ga \equiv dB_\ga +\go_\ga{}^\gb B_\gb\,,
$
and
\be e(l ,\lambda ,M)=\frac{1}{4}\lambda^2 -\half\frac{M^2}{l^2 -1}\,\qquad
( l\geq 2).  \ee
One can see that the freedom in an arbitrary parameter
$M$ is just the freedom of the relativistic field equations in the parameter
of mass.

Thus the equations (\ref{unads}) describe
a scalar field of an
arbitrary mass
in 2+1 dimensions. Now let us show how these equations can be
generated
with the
aid of the generalized oscillators (\ref{modosc}). To this end we
introduce the generating function
\bee
\label{gf}
C(q_\ga ,K|x) =
\sum^\infty_{n=0}\sum_{A=0,1}
\frac{1}{n!}C_{\ga_1 \ldots \ga_n }(x)(K)^A q^{\ga_1}\ldots q^{\ga_n}\,.
\eee

The relevant equations acquire then
the following simple form
\be
\label{coeq}
D C(q_\ga ,K|x) =\frac{1}{4i}\{ h^{\ga\gb}
q_\ga q_\gb , C(q,K |x)\}\,
\ee
(from now on we use the dimensionless units
with a unit AdS radius, $\lambda =1$).

To see that the integrability conditions
for (\ref{coeq})  reduce to the zero-curvature conditions for $sp(2)\oplus
sp(2)$ one observes that there is an automorphism of the AdS algebra which
changes a
sign of the AdS translations. This automorphism allows
one to introduce a ``twisted representation" of the AdS algebra
with the anticommutator instead of commutator in the translational
part of the AdS algebra. This twisted representation just leads to the
covariant constancy equations (\ref{coeq}).

Since the terms
in (\ref{coeq})
which depend on the background gravitational fields
only contain even combinations of the oscillators
$q_\ga$ the full system of equations decomposes into four independent
subsystems which can be singled out by virtue of the
projection operators
$P_\pm =\half (1\pm K)$ either in the boson or in the fermion sectors
(even (odd) functions
$C(q_\ga ,K|x)$ of
$q_\ga$
describe bosons (fermions)).
The explicit calculation which involves some reorderings of $q_\ga$
and rescalings of fields
then shows that the irreducible boson subsystems
 projected out by $P_\pm$ indeed reduce to the
equations of motion of the form (\ref{unads}) for
a massive scalar field
of mass $M^2 =\half\nu (\nu \mp 2)$. Remarkably, the same equations in the
fermion sector describe spin $\half$
fermion fields of the mass $M^2=\half\nu^2$.

An important achievement of the reformulation of the free field
equations in the form (\ref{coeq}) is that this form suggests
that the global HS symmetry algebra realized on the matter fields
of mass $M(\nu )$ is
$g=hs(2;\nu )\oplus hs(2;\nu )$ with the gauge fields (\ref{gau}).
To simplify the formulation it is convenient to introduce  two
Clifford variables
$\{\psi_i ,\psi_j \}=2\delta_{ij}$ $( i,j=1,2)$
instead of $\psi$.
One then introduces the full set of HS gauge fields as
$W_\nu (q_\ga ,K,\psi_{1,2} |x)$ and realizes
the gravitational fields as
\be
\label{gr}
W_\nu^{gr} =
\frac{1}{4i} (\go_\nu{}^{\ga\gb}+
h_\nu{}^{\ga\beta}\psi_1 )q_\ga q_\beta\,.
\ee
The generating function for
0-forms is
\bee
\label{maux}
C( q,K,\psi_{1,2} |x)=
C^{mat} ( q,K,\psi_1 |x)\psi_2
+
C^{aux} ( q,K,\psi_1 |x)\,.
\eee

Now let us consider the zero curvature equations
\bee
\label{zerc}
0\!=\!R\!\equiv \!dW (q,K,\psi |x )
+W (q,K,\psi |x)\wedge
W (q,K,\psi |x)
\eee
along with the covariant constancy conditions in the adjoint
representation of the HS algebra
\bee
0=dC (q,K,\psi |x)
+W (q,K,\psi |x)C (q,K,\psi |x)
-C (q,K,\psi |x)W (q,K,\psi |x)\,.
\eee
Due to the factor of $\psi_2$ in front of $C^{mat}$
the equations for $C^{mat}$ turn out to be
equivalent to the equations (\ref{coeq}) in the gauge in which
only the gravitational part (\ref{gr}) of the vacuum HS
gauge fields is non-vanishing. The fields $C^{aux}$
can be shown \cite{un} to be of a topological type
so that each irreducible subsystem in this sector can describe
at most a finite number of degrees of freedom and trivializes in
a topologically trivial situation. Thus the effect of introducing
a second Clifford element  consists of addition of some topological
fields.

\section{Non-linear dynamics}

To describe non-linear HS dynamics of
matter fields in 2+1 dimensions we start with a system of
equations which is very close to that introduced in \cite{eq}
for a particular case of massless matter fields. We
introduce three types of the generating functions
$dx^\nu W_\nu (z_\ga ,y_\gb ,K,\psi_i |x)$,
$ s_\gamma (z_\ga ,y_\gb ,K,\psi_i |x)$ and
$ B (z_\ga ,y_\gb ,K,\psi_i |x)$ which depend on the
space-time variables $x^\mu$ and auxiliary variables
 $(z_\ga ,\!y_\gb ,\!K,\!\psi_i )$ such that
 the two Clifford elements $\psi_i$
commute to all other variables, while the bosonic spinor
variables $z_\ga$ and $y_\gb$ commute to each other but anticommute
with $K\,,$ {\it i.e.}
$
\{K,z_\ga \}= \{K,y_\ga \}=0$, $ K^2=1\,.
$
Their physical content is as follows:
$dx^\nu W_\nu $ is the generating
function for HS gauge fields,
$ B $ contains physical matter degrees
of freedom along with some auxiliary variables,
and $ s_\gamma $ is entirely
auxiliary variable which  allows one to formulate the
full system of equations in a compact form.

This formulation
is based on the following star-product law which endows the space
of functions $f(z,y)$ with a structure of associative
algebra
\bee
\label{star}
(f*g)(z,y)\!=\!(2\pi )^{-2}\!\!\int d^2 ud^2 v\, f(z\!+\!u,y\!+\!u)
\times&\!\!\!\!g(z\!-\!v,y\!+\!v)\,\exp
i(u_\alpha v^\alpha )\,.
\eee
This product law provides a
particular symbol realization of the Heisenberg--Weyl algebra.
In particular one finds that
$
 [y_\ga ,y_\gb ]_* =-[z_\ga ,z_\gb ]_* =2i
\epsilon_{\ga\gb}\,.
$
The full system of equations has the form:
\bee
\label{zercur}
dW\!+\!W*\wedge W\!=\!0\,,\quad\!
ds_\ga \!+\!\tilde{W}* s_\ga \!-\!s_\ga * W\!=\!0\,,\quad\!
dB \!+\!{W}* B \!-\!B * W\!=\!0\,,
\eee
and
\bee
\label{cons}
\tilde{s}_\ga * s_\gb
-\tilde{s}_\gb * s_\ga=-2i\gee_{\ga\gb} (1+\kappa * B)\,,\qquad
\tilde{B}* s_\ga -s_\ga * B=0\,,
\eee
where
\be
\label{til}
\tilde{a}(z ,y ,K, \psi_i |x)
=a(z ,y ,-K, \psi_i |x) \quad\forall a
\ee
and $\kappa =K exp\,i(z_\ga y^\ga )$
 is a central element of the
algebra which has vanishing star commutators with $y_\ga$, $z_\ga$,
$K$ and $\psi_i$.

The system of equations (\ref{zercur}),(\ref{cons})
is explicitly invariant
under the general coordinate  transformations
and the
HS gauge transformations of the form
\bee
\label{trans}
\delta W=d\gee +W*\gee -\gee* W,\quad\!
\delta B=B*\gee -\gee* B,\quad\!
\delta s_\ga = s_\ga *\gee -\tilde{\gee}* s_\ga\,.
\eee
To elucidate its physical content
one has to analyze this system perturbatively near some
vacuum solution. In the massless case the appropriate vacuum solution
\cite{eq} is
\be
B_0 =0\,,\quad  s_0 {}_\ga =z_\ga\,,\quad   W_0 =\go (y,K,\psi_{1,2} )
\ee
with the vacuum gauge field $\go$ satisfying the zero curvature
condition $d\go +\go*\wedge\go=0\,.$
It can be shown along the lines of
\cite{eq} that the system of equations
(\ref{zercur}),(\ref{cons}) expanded near this vacuum solution
properly describes  dynamics
of massless matter fields
on the free field level and beyond.

The main result of this report consists \cite{pr2} of the observation that
the same system (\ref{zercur}), (\ref{cons}) expanded near another vacuum
solution
describes dynamics of matter fields with an arbitrary mass.
This is a solution with
\be
\label{vacnu}
B_0 =\nu \,,
\ee
 where $\nu$ is an
arbitrary constant. For a constant field $B_0$ only the first of the
equations (\ref{cons}) remains non-trivial. Remarkably it turns out to be
possible to find its explicit solution
\be
\label{s0}
{s}_0{}_\ga =z_\ga +
\nu (z_\ga -y_\ga )\int_0^1 dt te^{it(z_\beta y^\beta )}K
\ee
(it is not too difficult to check that (\ref{s0}) satisfies (\ref{cons})
 by a direct substitution).
Now let us turn to the equations (\ref{zercur}). The third of these
equations is trivially satisfied. The second one reads
\be
\label{W}
\tilde{W}* s_{0\ga} -s_{0\ga} * W=0\,,\qquad
\ee
where we have taken into account that $ds_{0\ga}=0$.
Eq.(\ref{W}) is
a complicated integral equation.
The key observation however is that it admits
the following two particular solutions: $W_0 =q_\ga$
$(\ga =1,2)$,
\be
q_\ga =y_\ga +
\nu K(z_\ga -y_\ga )\int_0^1 dt (1-t) e^{itz_\beta y^\beta }\,.
\ee
Taking into account that $*$-product is associative
it allows us to describe a general solution
of (\ref{W})
as an arbitrary
element
$
W_0 =\go (q_\ga , K, \psi_{1,2}|x)
$
 whose arguments
are treated as
some non-commutative elements of the star-product algebra.

To make contact
with the previous consideration it remains to check
by explicit computation that
the elements $q_\ga$ indeed obey
(\ref{modosc}). Thus, the vacuum solution with a constant field
(\ref{vacnu})
leads automatically to the deformed oscillator algebra with the
deformed oscillators realized as some functions of $z,y$ and $K$,
i.e. as elements of the tensor product of
two Heisenberg-Weyl algebras (equipped with the operator $K$).
Finally, it remains to observe that the first of the equations
(\ref{zercur})
reduces to the zero curvature equation which describes the AdS
background space. Since, as argued in the section 4,
$\nu$ governs the parameter of mass of matter fields we
arrive at the conclusion that a particular value of the parameter
of mass is determined by a vacuum value of the field $B$.

Next one can analyze the full system of equations perturbatively by
inserting the expansions of the form: $
W=W_0 +W_1 +\ldots\,,$ $ B=B_0 +B_1 +\ldots\,,$
$s_{\ga}=s_{0\ga} +s_{1\ga} +\ldots\,.$
In particular one can derive in the lowest orders that
\be
B_1 (z,y,K,\psi |x)=
C (q,K,\psi |x)\,,\quad\!
\ee
\be
W_1 (z,y,K,\psi |x)=
\go (q,K,\psi |x)+\Delta W_1 (C)\,,\qquad
s_{1\ga} =s_{1\ga} (C),
\ee
where $s_{1\ga} (C)$ and $\Delta W_1 (C)$ are some functionals
of the field $C$ which remains arbitrary and
has to be identified with generating function
(\ref{maux}). Inserting this back into
(\ref{zercur}) one obtains
the free field
equations for $C$
in the linearized approximation
from the third equation
and the equations of the form
$
d\go +\go*\wedge\go -J(\go, C^2)=0
$
from the first one where
$J(\go, C^2)$ is expected to describe HS currents (including
 the gravitational and spin - 1 ones).

\noindent
\section{Lorentz Covariance}

After it is argued that the system (\ref{zercur}), (\ref{cons})
describes properly HS dynamics in 2+1 dimensions let us explain
what a physical principle  fixes a particular form
of these equations. Remarkably, this is a simple and physically important
requirement that local Lorentz symmetry should be a particular
symmetry of the equations.

Let us consider the following element of the algebra
\be
\label{ltot}
L^{tot}_{\alpha \beta} =\frac{i}{4} \left (
\{z_\ga ,z_\gb \}_* -
\{y_\ga ,y_\gb \}_*
 \right )\,.
\ee
Infinitesimal local Lorentz transformations with a parameter
$\eta^{\ga\gb}$ are generated as
$
\delta Q(z,y)=[Q, \eta^{\ga\gb} L^{tot}_{\ga\gb}]_*\,.
$
Indeed these generators rotate properly the elementary spinor
generating elements of the algebra,
$
\delta z_\ga =\eta_{\ga\gb} z^\gb$,
$\delta y_\ga =\eta_{\ga\gb} y^\gb$,
and therefore induce principal $sp(2)$ transformations (automorphisms) of
the whole algebra.

Although the argument above proves explicit local Lorentz
invariance of the system of equations (\ref{zercur}), (\ref{cons}),
this symmetry is spontaneously broken due to
the first of the constraints (\ref{cons}). Indeed, since the right
hand side of this constraint has a non-vanishing vacuum value, $s_\ga$
itself must have a non-vanishing vacuum value (\ref{s0}). The question
therefore is whether there exists another local Lorentz symmetry which
rotates properly spinor indices of the dynamical fields leaving
invariant a vacuum solution. In fact the existence of such a
Lorentz symmetry in all orders in interactions is a highly non-trivial
property which fixes the constraints (\ref{cons}).

The point is that according to the analysis of the previous section
after the constraints (\ref{cons}) are solved the generating
functions for matter fields and gauge fields are described by
arbitrary functions of only one spinor variable $q_\ga$,
$C(q,K,\psi |x)$ and $\omega(q,K,\psi |x)$, respectively. In the
linearized approximation the Lorentz generators which rotate properly
$q_\ga$ are
$L_{\alpha \beta} =\frac{1}{4i}  \{q_\ga ,q_\gb \}_* $.
Thus  what we need is a proper generalization of these generators,
to all orders in interactions. The constraints
(\ref{cons}) indeed guarantee that such Lorentz generators
$l_{\ga\gb}$ can be constructed.

To see this we first observe that the constraints (\ref{cons}) give a
particular
realization of the deformed oscillator algebra
(\ref{modosc}). To this end
it is convenient to introduce a new
auxiliary generating element $\rho$
which has the properties
\be
\label{rho}
\{\rho ,K \}=0\,,\qquad \rho^2 =1\,.
\ee
Let us introduce a new variable $t_\ga =\rho s_\ga$. A role of $\rho$
is that it compensates the twiddle operation
(\ref{til}) in (\ref{zercur}) and
(\ref{cons}) so that the constraints (\ref{cons}) take the form
\bee
\label{const}
t_\ga * t_\gb
\!-\!t_\gb * t_\ga \!=\!-2i\gee_{\ga\gb} (1\!+\!\kappa * B),\quad\!\!
{B}* t_\ga \!-\!t_\ga * B\!=\!0,\quad\!\!
\kappa * t_\ga \!+\! t_\ga * \kappa\! =\!0,
\eee
where $\kappa =K exp\,i(z_\ga y^\ga )$ is a central element
of the original algebra which now anticommutes with $t_\ga$ due to
(\ref{rho}). To make contact with
(\ref{modosc}) one identifies $t_\ga$, $B$ and $\kappa$
with $iq_\ga$, $\nu$, and $K$, respectively
\footnote{Note that the vacuum solution $t_{0\ga}=\rho s_{0\ga}$
(\ref{s0}) therefore again describes an embedding of the deformed oscillator
algebra into a (equipped) direct product of the two Heisenberg-Weyl algebras.}
. As a consequence of the general property (\ref{q2com}) one concludes that
the elements
\be
\label{M}
M_{\alpha \beta} =\frac{i}{4}  \{t_\ga ,t_\gb \}_* =
\frac{i}{4}  (\tilde{s} _\ga * s_\gb  + \tilde{s} _\gb * s_\ga )
\ee
obey the Lorentz commutation relations and rotate properly $t_\ga$.
Now one can come back to the original $\rho$ - independent variables
arriving at the relations
\be
\label{Ms} \tilde{M}_{\ga \gb}*
s_{\ggg }- s_{\ggg }* M_{\ga \gb}
=\gee_{\ga\ggg}s_{\gb}+ \gee_{\gb\ggg}s_{\ga}
\,.
\ee

Let us now introduce the generators
\be
\label{Lph}
l_{\ga\gb} =
L^{tot}_{\ga\gb} -
M_{\ga\gb}\,.
\ee
{}From (\ref{cons}) it follows that
\be
\delta B = [B\,, \eta^{\ga\gb} l_{\ga\gb}]_* =
[ B\,,\eta^{\ga\gb} L^{tot}_{\ga\gb} ]_* \,,
\ee
i.e. $l_{\ga\gb}$ rotate properly physical fields like
$C(q,K,\psi |x)$ in all orders in interactions. Assuming that
$L^{tot}$ rotates properly $s_\ga$
one concludes
\footnote{
There is some gauge ambiguity in  the generic solution of the first of the
constraints (\ref{cons}) for $s_\ga$ in terms of $B$. The assumption
above is true when $s_\ga$ is reconstructed entirely in terms of $B$
without introducing any external constant spinors beyond those in
the vacuum solution $s_{0\ga}$.}
that
\be
\delta s_\ga =
s_\ga * \eta^{\gamma\gb} l_{\gamma\gb}  -
{\eta}^{\gamma\gb} \tilde{l}_{\gamma\gb}* s_\ga
  =\frac{\delta s_\alpha }{\delta B}{\delta B}
\ee
and that the gauge transformations induced by
$l_{\ga\gb}$ satisfy $sp(2)$ commutation relations.
Also
\be
\label{ll}
\delta W =
D (\eta^{\ga\gb} l_{\ga\gb} )
=d(\eta^{\ga\gb} )
l_{\ga\gb} +[W,
\eta^{\ga\gb} L^{tot}_{\ga\gb} ]
\ee
because $d(L^{tot}) =0$ while $D(M_{\ga\gb})=0$
as a consequence of the
second of the equations (\ref{zercur}).
{}From (\ref{ll}) one concludes
that the gauge field for a true local Lorentz symmetry is \be
W_L=\omega_L^{\ga\gb}l_{\ga\gb}
\ee
while all other gauge fields are rotated properly under the Lorentz
transformations.

Let us emphasize that the above analysis guarantees Lorentz symmetry in all
orders in interactions. Thus, it is the Lorentz symmetry principle which
fixes a form of the equations and enforces appearance of the deformed
oscillator algebra in the HS problem.

\noindent
\section{Concluding remarks}

\noindent

The proposed formulation of HS interactions admits interpretation
of the parameter of mass as a module of the space of vacuum solutions,
i.e. the same equations describe HS interactions of
massive multiplets with different masses depending on a chosen vacuum
solution. As a result different global HS
symmetries of the linearized matter multiplets are
different stability subgroups of the full HS
symmetry, which leave invariant vacuum solutions.
These global HS symmetries turn out to be pairwise non-isomorphic for
different values of the parameter of mass.
It is worth mentioning that the model under consideration (eq.(\ref{coeq}))
possesses $N=2$ supersymmetry $osp(2;2)$ with the generators (\ref{N2}).
The constraints have a form of the deformed oscillator algebra as a
consequence of the requirement that the equations of motion of matter fields
interacting with HS fields must possess local Lorentz symmetry which
is guaranteed by the properties (\ref{q1com}) and (\ref{q2com}).

\vskip0.1cm
The research described in this report
 was supported in part by the
Russian Foundation for Basic Research, Grant No.96-01-01144 and
by the European Community
Commission under the contract INTAS, Grant No.93-633{\it -ext}.

\end{document}